\documentclass[preprint,prd,nofootinbib,tightenlines,amsmath,amssymb]{revtex4}
\usepackage{enumerate}
\usepackage{multirow}
\usepackage[utf8]{inputenc}
\usepackage{anysize}
\usepackage{textcomp}
\usepackage{epsfig}
\usepackage{graphics,color} 
\usepackage{verbatim}
\usepackage{afterpage}
\usepackage{amsmath}    
\usepackage{soul}
\usepackage{xcolor,cancel}
\DeclareUnicodeCharacter{00A0}{ }

\usepackage{blindtext}

\catcode`\@=11
\def\sla@#1#2#3#4#5{{%
 \setbox\z@\hbox{$\m@th#4#5$}%
 \setbox\tw@\hbox{$\m@th#4#1$}%
 \dimen4\wd\ifdim\wd\z@<\wd\tw@\tw@\else\z@\fi
 \dimen@\ht\tw@
 \advance\dimen@-\dp\tw@ \advance\dimen@-\ht\z@
 \advance\dimen@\dp\z@
 \divide\dimen@\tw@ \advance\dimen@-#3\ht\tw@
 \advance\dimen@-#3\dp\tw@ \dimen@ii#2\wd\z@
 \raise-\dimen@\hbox to\dimen4{%
 \hss\kern\dimen@ii\box\tw@\kern-\dimen@ii\hss}%
 \llap{\hbox to\dimen4{\hss\box\z@\hss}}}}

\def\cpto{\mathrel {\vcenter {\baselineskip 0pt \kern 0pt
    \hbox{$H_{r.f.}$} \kern 0pt \hbox{$\longrightarrow$} }}}
\def\slashed#1{%
 \expandafter\ifx\csname sla@\string#1\endcsname\relax
{\mathpalette{\sla@/00}{#1}}
\fi}
\def\declareslashed#1#2#3#4#5{%
 \expandafter\def\csname sla@\string#5\endcsname{%
#1{\mathpalette{\sla@{#2}{#3}{#4}}{#5}}}}
 \catcode`\@=12
\declareslashed{}{/}{.08}{0}{D}
 \declareslashed{}{/}{.1}{0}{A}
 \declareslashed{}{/}{0}{-.05}{k}
 \declareslashed{}{/}{.1}{0}{\partial}
 \declareslashed{}{\not}{-.6}{0}{f}

\def\lsim{\mathrel {\vcenter {\baselineskip 0pt \kern 0pt
    \hbox{$<$} \kern 0pt \hbox{$\sim$} }}}
\def\gsim{\mathrel {\vcenter {\baselineskip 0pt \kern 0pt
    \hbox{$>$} \kern 0pt \hbox{$\sim$} }}}

\newcommand{\bea}{\begin{eqnarray}}
\newcommand{\eea}{\end{eqnarray}}

\begin{document}

\baselineskip=15pt
\preprint{NCTS-PH1706}

\title{Dark Photon Search at A Circular $e^+e^-$ Collider}

\author{Min He$^1$\footnote{Electronic address: hemin\_sjtu@163.com}, Xiao-Gang He$^{2,3,1}$\footnote{Electronic address: hexg@phys.ntu.edu.tw}, Cheng-Kai Huang$^{2}$\footnote{Electronic address: r01222045@ntu.edu.tw }}
\affiliation{
$^{1}$INPAC,Department of Physics and Astronomy, Shanghai Jiao Tong University, Shanghai.\\
$^{2}$Department of Physics, National Taiwan University, Taipei. \\
$^{3}$Physics Division, National Center for Theoretical Sciences, Hsinchu, Taiwan 30013.
}

\date{\today}

\vskip 1cm
\begin{abstract}

One of the interesting portals linking a dark sector and the standard model (SM)  is the kinetic mixing between the SM  $U(1)_Y$ field with a new dark photon $A'$ from a $U(1)_{A'}$ gauge interaction. Stringent limits have been obtained for the kinetic mixing parameter $\epsilon$ through various processes. In this work, we study the possibility of searching for a dark photon interaction at a circular $e^+e^-$ collider through the process $e^+ e^-\to \gamma A^{\prime *} \to \gamma \mu^+\mu^-$. We find that the constraint on $\epsilon^2$ for dark photon mass in the few tens of GeV range, assuming that the $\mu^+\mu^-$ invariant mass can be measured to an accuracy of $0.5\%m_{A'}$,
can be better than $3\times 10^{-6}$ for the proposed CEPC with a ten-year running at 3$\sigma$ (statistic) level, and better than $2\times 10^{-6}$ for FCC-ee with even just one-year running at $\sqrt{s} = 240$ GeV, better than the LHC and other facilities can do in a similar dark photon mass range. For FCC-ee, running at $\sqrt{s}=160$ GeV, the constraint can be even better.

\end{abstract}

\pacs{PACS numbers: }

\maketitle

\noindent
{\bf Introduction}

There may be a dark sector  which only interacts indirectly with the standard model (SM) particles through some sort of portal. One of the interesting portal possibilities is the dark photon $A'$ of an additional $U(1)_{A'}$ gauge symmetry interacting with SM particles resulted from kinetic mixing with the SM $U(1)_Y$ gauge boson\cite{kinetic-mixing,kinetic-mixing1}. There are many interesting consequences if such a dark photon exists\cite{kinetic-mixing2,kinetic-mixing3}. Great efforts have been made to search for a dark photon through various processes and stringent limits have been obtained for the kinetic mixing parameter for a given dark photon mass $m_{A'}$\cite{limits,limits1,limits2}. Most of the constraints on dark photon kinetic mixing are for dark photon with a low mass (less than 10 GeV or so) from various low energy facilities and rare decays of known particles.  There is no convincing theoretical argument that the dark photon must have a low mass. It would be better to have experimental data to tell whether larger dark photon mass is allowed. There are less studies of constraints on dark photon with a larger mass. LHC may provide some important information\cite{limits1}. LHCb can provide stringent constraint on the kinetic mixing for dark photon mass larger than 10 GeV. It has been shown that the ATLAS and CMS may provide even better constraint at dark photon mass around 40 to 50 GeV by studying $pp \to X \mu^+\mu^-$. The LHC can provide higher energy to probe larger dark photon mass. However, if analysis can be carried out at a high energy $e^+ e^-$ colliders , such as the CEPC and FCC-ee, the background and signal may be easier to separate and provide better information. A possible process is $e^+e^-\to \gamma \mu^+\mu^-$. The final states $\gamma$ and $\mu^+\mu^-$ in this case can be more easily studied compared with the $X\mu^+\mu^-$ final states in the $pp$ collision case. Better constraint may be possible. 
In this work, we study the possibility to search for dark photon effects at a circular $e^+e^-$ collider through the process $e^+ e^-\to \gamma A^{\prime *} \to \gamma \mu^+\mu^-$. 

A dark photon $A'$ from an extra $U(1)_{A'}$ gauge interaction, which does not couple to SM fields directly, can indirectly interact through a renormalizable kinetic mixing term $F'_{\mu\nu} B^{\mu\nu}$ with SM particles. Here $F'_{\mu\nu}=\partial_\mu A^{\prime}_\nu-\partial_\nu A^{\prime}_\mu$ and $B_{\mu\nu} = \partial_\mu B_\nu-\partial_\nu B_\mu$, and $A'$ and $B$ are the $U(1)_{A'}$ and $U(1)_Y$ gauge fields, respectively. With kinetic mixing, the renormalizable terms involving these two $U(1)$ gauge fields are given by\cite{kinetic-mixing}
\begin{eqnarray}
L_{\mbox{kinetic}} = -{1\over 4} B_{0}^{\mu\nu}B_{0,\mu\nu} -{1\over 2} \sigma F'_{0, \mu\nu}B_0^{\mu\nu} - {1\over 4} F'_{0,\mu\nu}F_0^{\prime \mu\nu}\;.
\end{eqnarray}

The $U(1)_Y$ gauge field $B_0$ is a linear combination of the photon $A_0$ and the $Z_0$ boson fields, $B_0 = c_W A_0 - s_W Z_0$ with $c_W = \cos\theta_W$ and $s_W = \sin\theta_W$.  $\theta_W$ is the weak interaction Weinberg angle.
To have the above Lagrangian in the canonical form, that is, there are no crossing terms, one needs to redefine the fields. Letting the redefined fields to be $A$, $Z$ and $A'$, we have\cite{kinetic-mixing}
\begin{eqnarray}
\left ( \begin{array}{c}
A_0\\
Z_0\\
A'_0
\end{array}
\right ) = \left ( \begin{array}{ccc}
 1&0&-{c_W \sigma\over \sqrt{1-\sigma^2}}\\
 0&1&{s_W \sigma\over \sqrt{1-\sigma^2}}\\
 0&0&{1\over \sqrt{1-\sigma^2}}
 \end{array}
 \right )
 \left (\begin{array}{c}
 A\\Z\\A'
 \end{array}
 \right )\;.
 \end{eqnarray}
 
The interaction of $A$, $Z$ and $A'$ with SM currents, to the first order in $\sigma$ is given by
\begin{eqnarray}
J^\mu_{em} (A_\mu - c_W \sigma A'_\mu) + J^\mu_{Z} (Z_\mu + s_W\sigma A'_\mu) + J^\mu_{D} A'_\mu \;,
\end{eqnarray}
where $J^\mu_{em}$, $J^\mu_Z$ are the SM electromagnetic and $Z$ boson interaction currents, respectively. 
$J^\mu_D$ is the dark current in the dark sector.

After the electroweak symmetry breaking, the $Z_0$ boson obtains a non-zero mass $m_Z$. Depending on how the $U(1)_{A'}$ symmetry is broken, $A'_0$ boson can receive a non-zero mass which may or may not cause mixing with $Z_0$. If one introduces a SM singlet $S$ with a non-trivial $U(1)_{A'}$ quantum number $s_{A'}$ to break the symmetry, $A'_0$ boson
will receive  a mass $m_{A'} = g_{A'}s_{A'}v_s/\sqrt{2}$. Here $g_{A'}$ is the $U(1)_{A'}$ gauge coupling constant and $v_s/\sqrt{2}$ is the vacuum expectation value $\langle S \rangle$ of $S$ field, In the $Z$ and $A'$ basis, they mix with each other with the mixing matrix  given by
\begin{eqnarray}
\left ( \begin{array}{cc}
 m^2_Z&{\sigma s_W\over \sqrt{1-\sigma^2}}m^2_Z\\
 {\sigma s_W\over \sqrt{1-\sigma^2}}m^2_Z&{1\over 1-\sigma^2}m^2_{A'} + {s^2_W \sigma^2\over 1-\sigma^2} m^2_Z
 \end{array}
 \right )
\end{eqnarray}
For a small $\sigma$, in the mass eigenstate bases, $A'$ only interacts with $J_Z^\mu$ at the second order in 
$\sigma$.

In general if the vacuum expectation value of the scalar which breaks $U(1)_{A'}$ also carries $U(1)_Y$ charge\cite{kinetic-mixing1}, 
there is an additional mixing parameter between $Z$ and $A'$ which can lead to an interaction of $A'$ with $J_Z$. But, the interaction of $A'$ with $J_{em}$ remains the same\cite{kinetic-mixing1}. 
In our later discussions, we will take the simple case that $S$ does not carry any $U(1)_{Y}$ charge and concentrate on the study of 
the search of dark photon $A'$ at a circular $e^+e^-$ collider through $e^+e^- \to  \gamma \mu^+\mu^-$. 
To have a similar notation compared with that used by many in the literature, we use the notation $-c_W\sigma = \epsilon$. Besides 
$A'$ contribution to $e^+e^- \to  \gamma \mu^+\mu^-$ through intermediate $A'$, there are also SM contributions from intermediate $A$ and $Z$ interactions. The effective Lagrangian concerning photon and dark photon interaction with SM currents to be used is in the following form
\begin{eqnarray}
L_{\mbox{int}} =J^\mu_{em} A_\mu + J^\mu_Z Z_\mu + \epsilon J^\mu_{em} A'_\mu\;.
\end{eqnarray}

At low energy, way below $Z$ boson mass, the contribution from intermediate $Z$ boson is small. When the collision energy is large compared with the $Z$ boson mass, the interaction term $J^\mu_{Z} Z_\mu$ may also be important. For searching for dark photon effects through $e^+e^- \to \gamma A^{\prime*} \to \gamma \mu^+\mu^-$, the SM contributions, $e^+e^- \to \gamma (\gamma^*, Z^*) \to \gamma \mu^+\mu^-$ become the background.  
We find that the constraint on $\epsilon^2$ for dark photon mass in the few tens of GeV range, assuming that the $\mu^+\mu^-$ invariant mass can be measured to an accuracy of $0.5\%m_{A'}$,
can be better than $3\times 10^{-6}$ for the proposed CEPC with a ten-year running at 3$\sigma$ (statistic) level, and better than $2\times 10^{-6}$ for FCC-ee with even just one-year running at $\sqrt{s} = 240$ GeV, better than the LHC and other facilities can do in a similar dark photon mass. For FCC-ee, running at $\sqrt{s}=160$ GeV, the constraint can be even better.
\\

\noindent
{\bf The $e^+ e^- \to \gamma (\gamma^*, Z^*, A^{\prime *}) \to \gamma \mu^+\mu^-$ processes}

The Feynman diagram for $e^+e^-\to \gamma (\gamma^*, Z^*,  A^{\prime *})\to \gamma \mu^+\mu^-$ are shown in Fig. 1. The contributions from intermediate $A'$ state are suppressed by a factor of $\epsilon^4$ which can naively be thought to be negligible compared with 
intermediate $\gamma, Z$ contributions\cite{calculations}. However, since the final invariant mass of the $\mu^+\mu^-$ pair is not fixed, the value can hit the $A'$ mass pole, the cross section $\sigma^{m_{\mu\mu}}_{\gamma A'}(e^+e^-\to \gamma A^{\prime *} \to \gamma \mu^+\mu^-)$ at the $A'$ mass pole can be large, even larger than the SM cross section $\sigma^{m_{\mu\mu}}_{\gamma(\gamma, Z)}(e^+e^-\to \gamma(\gamma^*, Z^*)\to \gamma \mu^+\mu^-)$, with the $\mu^+\mu^-$ invariant mass $s_3 = (k_1+k_2)^2 = m^2_{\mu\mu}$ close to $m^2_{A'}$, that is, $m_{\mu\mu}$ in the range of $m_{A'} - \sigma_{\mu\mu} \sim m_{A'}+ \sigma_{\mu\mu}$ with $\sigma_{\mu\mu}$ to be much smaller than $m_{A'}$. The subscript $(\gamma, Z)$ indicates contributions from both intermediate $\gamma$ and $Z$. Measurement at that region with small enough $\sigma_{\mu\mu}$ can provide information about the dark photon interaction. In the following we explain how this can be done. 

To this end 
we defined below two measurable quantities,
\begin{eqnarray}
&&\sigma^{m_{\mu\mu}}_{\gamma A'} =\int^{(m_{A'}+\sigma_{\mu\mu})^2}_{(m_{A'}-\sigma_{\mu\mu})^2} (d\sigma_{\gamma A'}/ds_3) ds_3\nonumber\\
&&\sigma^{m_{\mu\mu}}_{\gamma(\gamma, Z)} = \int^{(m_{A'}+\sigma_{\mu\mu})^2}_{(m_{A'}-\sigma_{\mu\mu})^2} (d\sigma_{\gamma (\gamma,Z)}/ds_3) ds_3\;.
\end{eqnarray}

\begin{figure}[!hbt]
\centering
\includegraphics{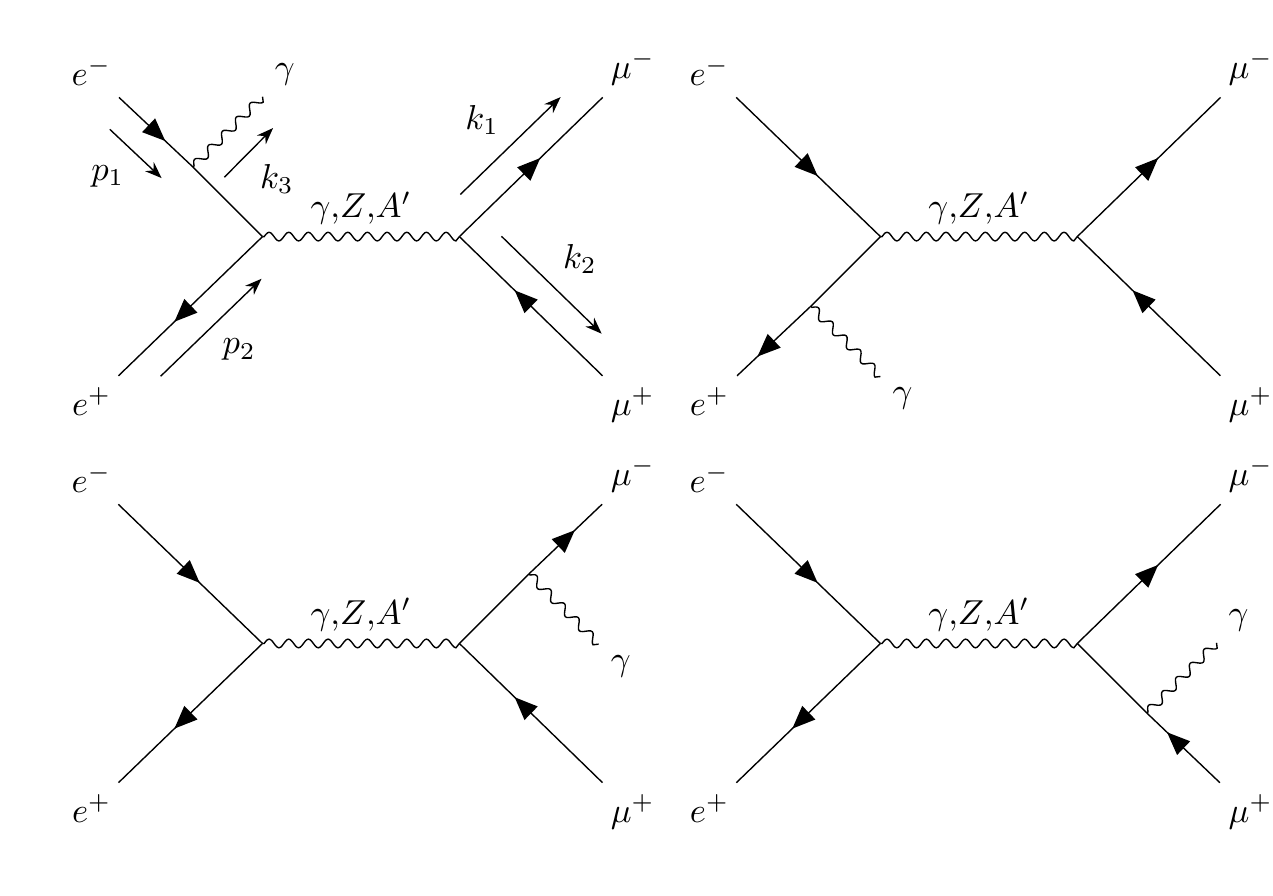}
\caption{Feynman diagrams for $e^- e^+ \to \gamma \mu^- \mu^+$ process.}
\label{feynman-diagram}
\end{figure}

Evaluating the Feynman diagrams with intermediate $\gamma$, $Z$ and $A^{\prime}$ shown in Fig. 1, we obtain
\begin{eqnarray}
{d\sigma_{\gamma(\gamma,Z)}\over ds_3} &=& {4 \alpha^3_{em} (s^2+s_3^2)\over 3 s^3 s_3(s-s_3)}\left (
s (\ln (s/m_e^2 ) -1) + s_3 ( \ln (s_3/ m_\mu^2 ) -1) \right )\nonumber\\
&&+{\alpha^3_{em} (8\sin^4\theta_W - 4 \sin^2\theta_W+1)^2\over 48 \sin^4\theta_W \cos^4\theta_W} 
{s^2+s^2_3\over s^2(s-s_3)}\nonumber\\
&&\times \left ({s_3 (\ln (s/m^2_e) -1 )\over (s_3-m^2_Z)^2} + {s (\ln (s_3/m^2_\mu ) -1)\over (s-m^2_Z)^2}\right )\nonumber\\
&&-{\alpha^3_{em} (1-4\sin^2\theta_W)^2\over 16 \sin^4\theta_W\cos^4\theta_W}{s+s_3\over s^2(s-s_3)}{ss_3\over (s-m_Z^2)(s_3 - m^2_Z)}\nonumber\\
&&+{\alpha^3_{em} (1-4\sin^2\theta_W)^2\over 6 \sin^2\theta_W \cos^2\theta_W}{s^2+s_3^2\over s^2(s-s_3)}
\left ({\ln(s/m^2_e)-1\over s_3-m^2_Z}  +{\ln(s_3/m^2_\mu) - 1\over s-m^2_Z}\right )\nonumber\\
&&-{\alpha^3_{em}\over 4 \sin^2\theta_W \cos^2\theta_W} {s+s_3\over s^2(s-s_3)} \left ({s_3\over s_3-m^2_Z} + {s\over s-m^2_Z}\right)\;;
\nonumber\\
\\
{d\sigma_{\gamma A'}\over ds_3} &\approx& {4 \alpha^3_{em}\epsilon^4 s_3 (s^2+s_3^2)\over 3 s^2(s-s_3)((s_3-m^2_{A'})^2 + \Gamma_{A'}^2 m^2_{A'})}\left ( \ln(s/ m_e^2) -1\right )\nonumber\\
&\approx& {4 \alpha^3_{em}\epsilon^4 s_3 (s^2+s_3^2)\over 3 s^2(s-s_3)}{\pi\over \Gamma_{A'} m_{A'}}\delta(s_3 - m^2_{A'})\left (\ln (s/ m_e^2 ) -1\right )\;,\nonumber
\end{eqnarray}
where $\alpha_{em}$ is the fine structure constant, and $s= (p_1+p_2)^2$.

For $d\sigma_{\gamma A'}/ds_3$, we have kept the leading contribution which is, in the narrow width approximation, proportional to $\delta (s_3 - m^2_{A'})$. $\Gamma_{A'}$ is the decay width of the dark photon
\begin{eqnarray}
\Gamma_{A'} = \sum_f \Gamma(A' \to \bar f f)\;,\;\;\Gamma(A' \to \bar f f ) = \epsilon^2 {Q_f^2 \alpha_{em} m_{A'} \over 3}(1+{2m^2_f \over m^2_{A'}}) \sqrt{1-{4m^2_f \over m^2_{A'}}}\;.
\end{eqnarray}
The summation above is to sum over fermion pairs in the SM  with mass $m_f < m_{A'}/2$. We find that in the range of a few tens of GeV up to 60 GeV or so (significantly below the $Z$ pole), the process may be able to provide stringent constraints on $\epsilon^2$, we will work in this region. In this case the summation of $f$ will need to sum over u, d, s, c, b, e, $\mu$ and $\tau$. With lower $A'$ mass, one should be careful to only sum over states which are below the threshold. 

Multiplying the integrated luminosity $I = \mbox{time} \times \mbox{luminosity}$, one obtains the event numbers for the SM contributions $N_{\gamma(\gamma, Z)}$ and dark photon contribution $N_{\gamma A'}$ respectively
\begin{eqnarray}
N_{\gamma(\gamma,Z)} = \sigma^{m_{\mu\mu}}_{\gamma(\gamma,Z)} I \;,\;\;\;\;N_{\gamma A'} = \sigma^{m_{\mu\mu}}_{\gamma A'} I\;.
\end{eqnarray}

With information from event numbers which can be obtained, one can analyze the sensitivity for a given experiment and then obtain the constraints on $\epsilon^2$ as a function of the dark photon mass $m_{A'}$.
\\

\noindent
{\bf Sensitivities for CEPC and FCC-ee circular colliders}

The sensitivity on the mixing parameter $\epsilon$ at an $e^+e^-$ collider depends on how well one can control over the $\mu^+\mu^-$ invariant mass measurement, that is how small one can accurately control $\sigma_{\mu\mu}$. We will take a similar accuracy as what can be done at the LHC\cite{limits2}, assuming $\sigma_{\mu\mu} = 0.5\% m_{A'}$.  In the study of searching for dark photon in Ref.\cite{limits2}, for SM contribution, only intermediate $\gamma$ contribution was considered. When include intermediate $Z$ contribution, the detailed values for the sensitivity will be changed. The statistic sensitivity depends on how one can measure the background events from SM contribution. We take $\sqrt{N_{\gamma(\gamma,Z)}}$ as the statistic sensitivity for SM background. We then obtain the signal $S$ to background error $\sqrt{N_{\gamma(\gamma,Z)}}$ ratio $\chi = N_{\gamma A'}/\sqrt{N_{\gamma(\gamma,Z)}}$ as the indicator how well one can obtain constraints on $\epsilon^2$ and $m_{A'}$. As long as statistic errors are concerned, the value of $\chi$ corresponds to the number of $\sigma$. There may be other background coming from other processes and also systematic errors which required a more involved analysis with full knowledge of the detectors. Here we will only consider statistic error discussed above. We will obtain the event numbers 
using the benchmark luminosities planed for the CEPC\cite{cepc}: $2 \times 10^{34} \mbox{cm}^{-2} \mbox{s}^{-1}$ at $\sqrt{s}$ = 240 GeV, and 
FCC-ee\cite{FCC-ee}: $1.5\times 10^{36} \mbox{cm}^{-2}\mbox{s}^{-1}$, $3.5\times 10^{35} \mbox{cm}^{-2}\mbox{s}^{-1}$, and $8.4\times 10^{34} \mbox{cm}^{-2}\mbox{s}^{-1}$ at $\sqrt{s}$ equal to 160 GeV, 240 GeV and 350 GeV, respectively. The results on the cross section, event numbers per year and the sensitivity for the mixing parameter $\epsilon$ are shown in Figs. 2, 3, 4 and 5.

The cross sections for various energies relevant to CEPC and FCC-ee are shown in Fig. 2. The cross sections for SM contributions are above $5\times 10^{-40} \mbox{cm}^2$ for $\sqrt{s}$ in the whole range covered by CEPC and FCC-ee. 
The SM background have two contributions, the intermediate $\gamma$ and $Z$ contributions. In the parameter space we are considering the intermediate $\gamma$ contribution dominates. Numerically we find that the intermediate $Z$ contribution is less than 10\% of the intermediate $\gamma$ contribution in the parameter spaces discussed above. If the machine is running at the $Z$ pole, the intermediate $Z$ contribution become important. Also when the dark photon mass is close to the $Z$ mass, the dark photon contribution also become important. For these reasons, we have limited the dark photon mass to be significantly away from the the $Z$ mass and also have chosen machine energies to be away from the $Z$ boson mass. 

Multiplying the luminosity of each machine at different energies on the cross section shown in Fig. 2, we obtain the one-year running event numbers shown in Fig. 3. We see that even for the lowest case, the CEPC case, the event number per year for SM contribution can be more than 1000. This provides a large enough number for analysis with some accuracy.

The sensitivity of $\epsilon^2$ as a function of dark photon mass $m_{A'}$ for a given $\chi$ are shown in Figs. 4 and 5.  
We plot $\epsilon^2$ as a function of $m_{A'}$ for various values $\chi$ for different machines and running times.

For CEPC, in the range of 10 GeV to 60 GeV for $m_{A'}$, the one-year running sensitivity for $\epsilon^2$ can reach $9\times 10^{-6}$ at 3$\sigma$ level. If one just wants an 1$\sigma$ limit, the sensitivity for $\epsilon^2$ can reach $3\times 10^{-6}$. With ten-year running time, the sensitivity can lower a factor of $\sqrt{10}$. These limits are better than ATLAS and CMS at the LHC can reach\cite{limits2}. In the range of 1 GeV to 10 GeV for $m_{A'}$, the sensitivity for $\epsilon^2$ is better because the width $\Gamma_{A'}$ is smaller. However, other processes, such as Belle II experiment, can give better constraints\cite{limits2}. At 10 GeV to 20 GeV $m_{A'}$ mass range, LHCb may give a slightly better constraint\cite{limits2}.

For FCC-ee, since the luminosity at $\sqrt{s} =240$ GeV is higher than that for CEPC, the sensitivity can be much better. The one-year running sensitivity for $\epsilon^2$
at $\sqrt{s} = 160$ GeV, can reach $0.7\times 10^{-6}$ at 3$\sigma$ level. The 1$\sigma$ limit can reach $2.4\times 10^{-7}$.
At $\sqrt{s}=240$ GeV, the one-year running sensitivity for $\epsilon^2$ can reach $2.2\times 10^{-6}$ at 3$\sigma$ level. The 1$\sigma$ limit can reach $0.72\times 10^{-7}$. At $\sqrt{s}=350$ GeV, the one-year running sensitivity for $\epsilon^2$ can reach $6.5\times 10^{-6}$ at 3$\sigma$ level. The 1$\sigma$ limit can reach $2\times 10^{-6}$.
\\

\begin{figure}[!hbt]
\centering
\begin{tabular}{cc}
\includegraphics[width=0.45\textwidth]{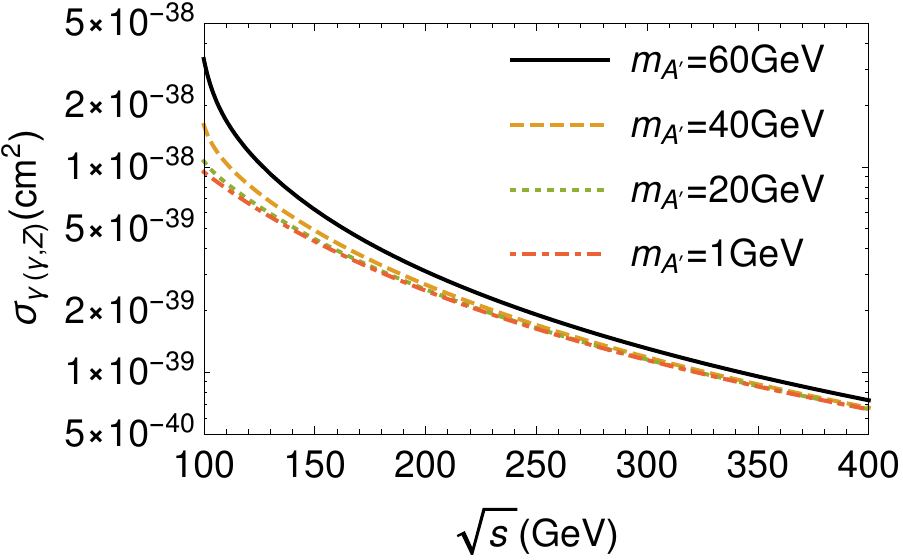}
\vspace*{1em}
\includegraphics[width=0.45\textwidth]{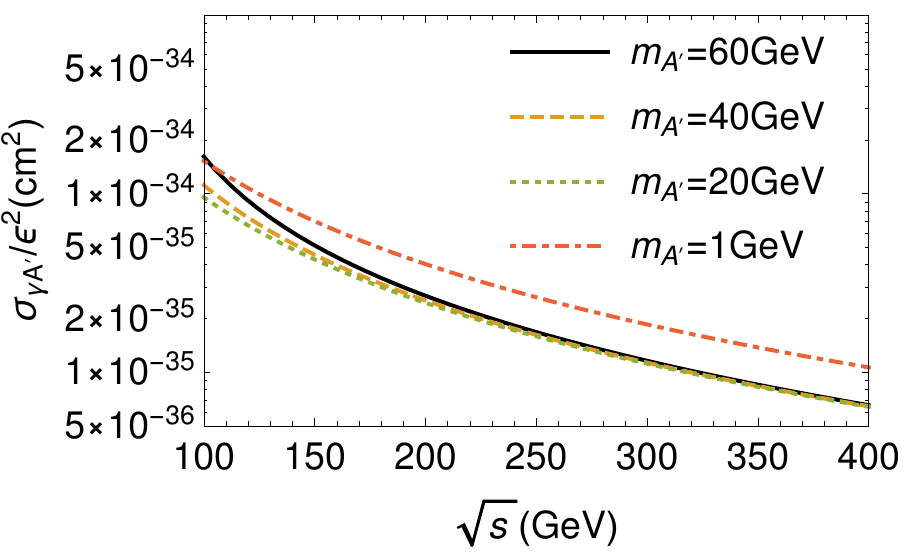}
\end{tabular}
\caption{Cross sections $\sigma^{m_{\mu\mu}}_{\gamma(\gamma,Z)}$ and $\sigma^{m_{\mu\mu}}_{\gamma A'}$ as functions of $\sqrt{s}$.
$m_{A'} = 1, 20, 40\; \mbox{and}\; 60$ GeV. The $m_{A'}$ dependence of $\sigma^{m_{\mu\mu}}_{\gamma(\gamma,Z)}$ is due to integration ranges depend on $m_{A'}$.}
\label{diff-cross-sec-fig}
\end{figure}

\begin{figure}[!hbt]
\centering
\begin{tabular}{cc}
\includegraphics[width=0.45\textwidth]{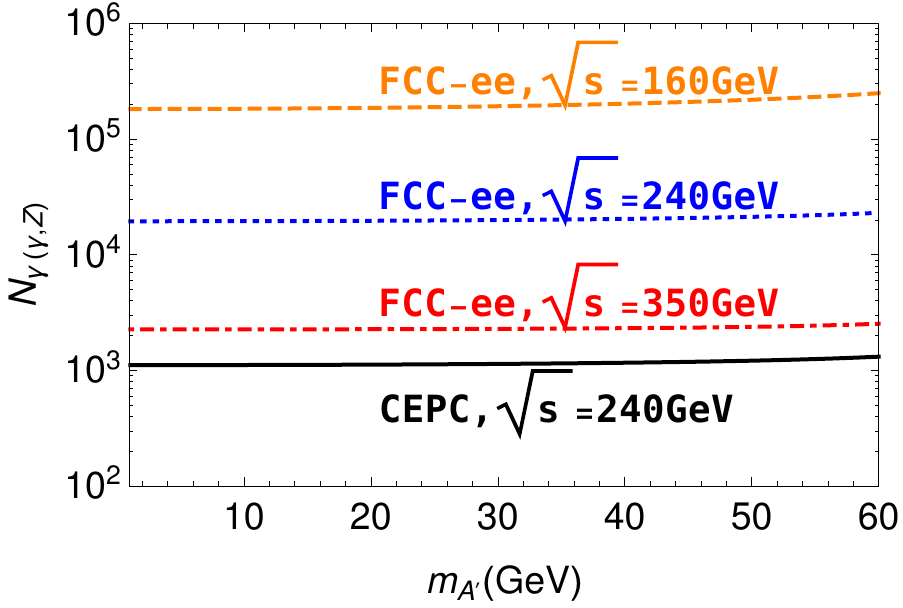}
\vspace*{1em}
\includegraphics[width=0.45\textwidth]{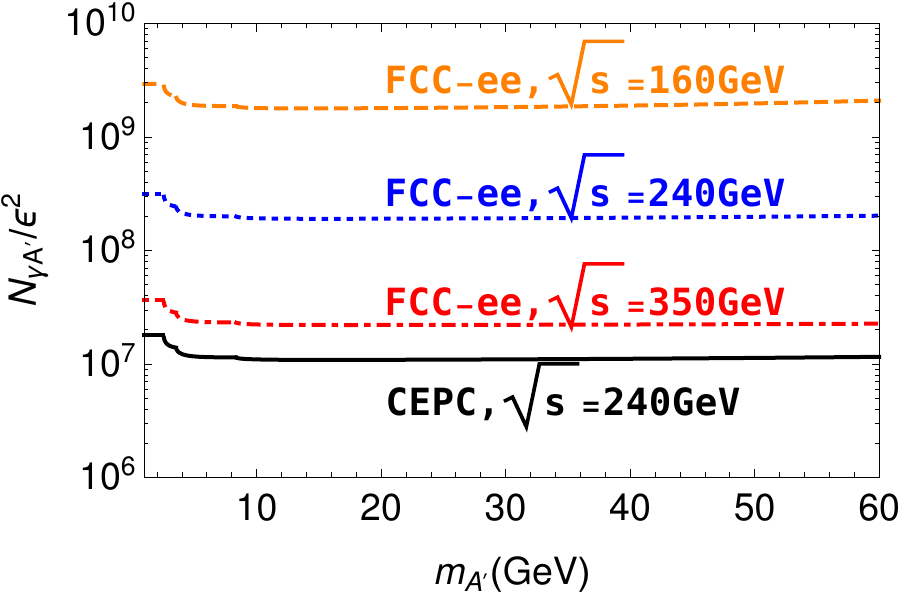}
\end{tabular}
\caption{$N_{\gamma (\gamma,Z)}$ and $N_{\gamma A'}/\epsilon^2$ as functions of $m_{A'}$ for CEPC ($\sqrt{s } = 240$ GeV) and FCC-ee ($\sqrt{s} = 160,\;240,\;350$ GeV) for one-year running. }
\label{event-N}
\end{figure}

\begin{figure}[!hbt]
\centering
\includegraphics[width=0.45\textwidth]{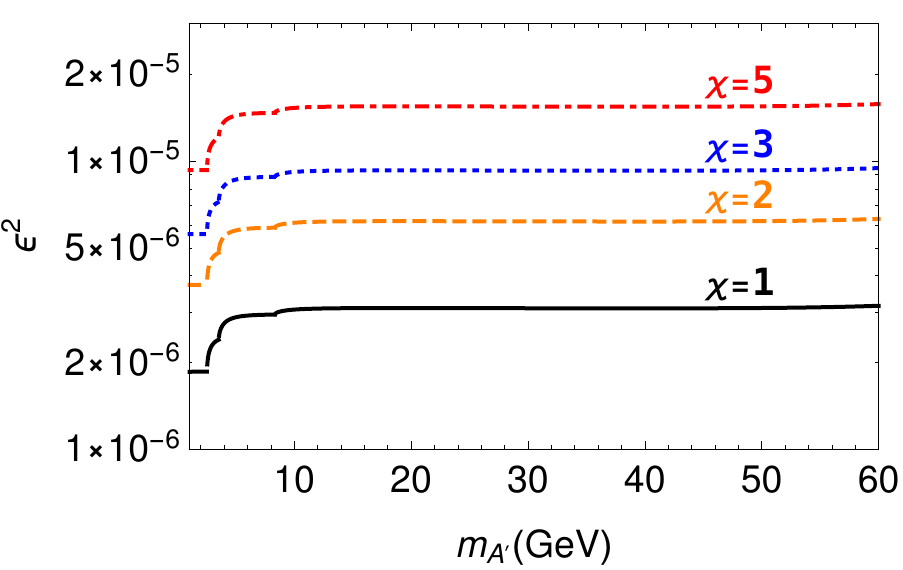}
\vspace*{1em}
\includegraphics[width=0.45\textwidth]{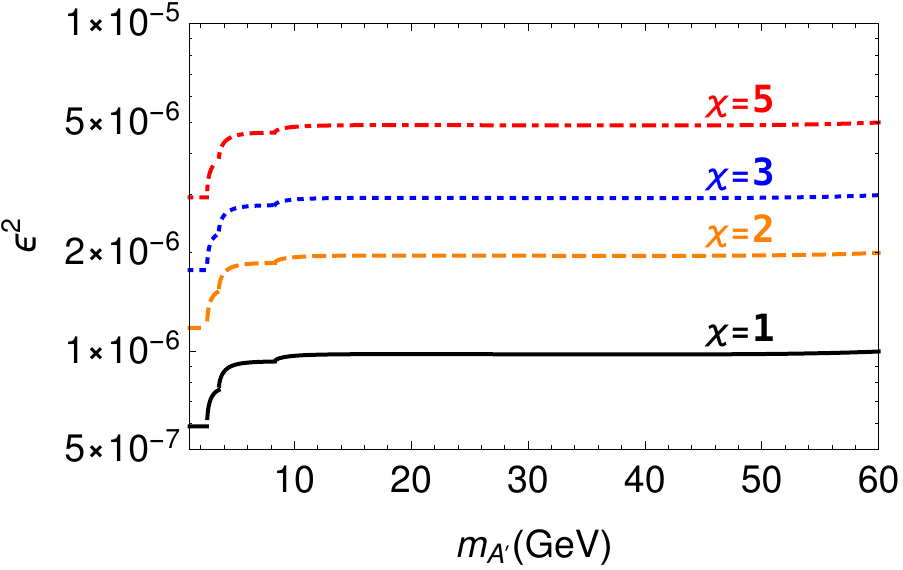}
\caption{Sensitivity on $\epsilon^2$ as functions of dark photon mass $m_{A^{\prime}}$ for a given $\chi$ for CEPC. The left figure is for one-year running and the right figure is for ten-year running.}
\label{fig4}
\end{figure}

\begin{figure}[!hbt]
\centering
\includegraphics[width=0.3\textwidth]{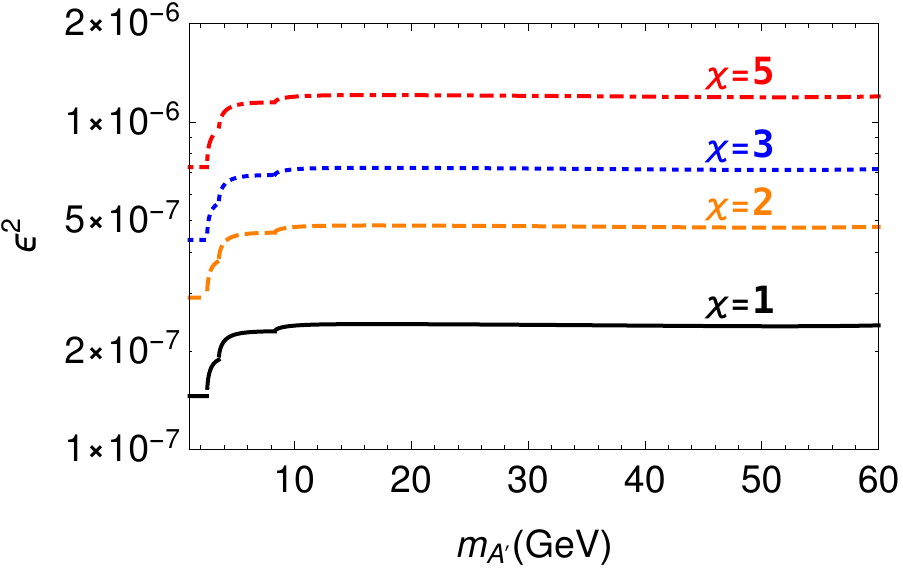}
\vspace*{1em}
\includegraphics[width=0.3\textwidth]{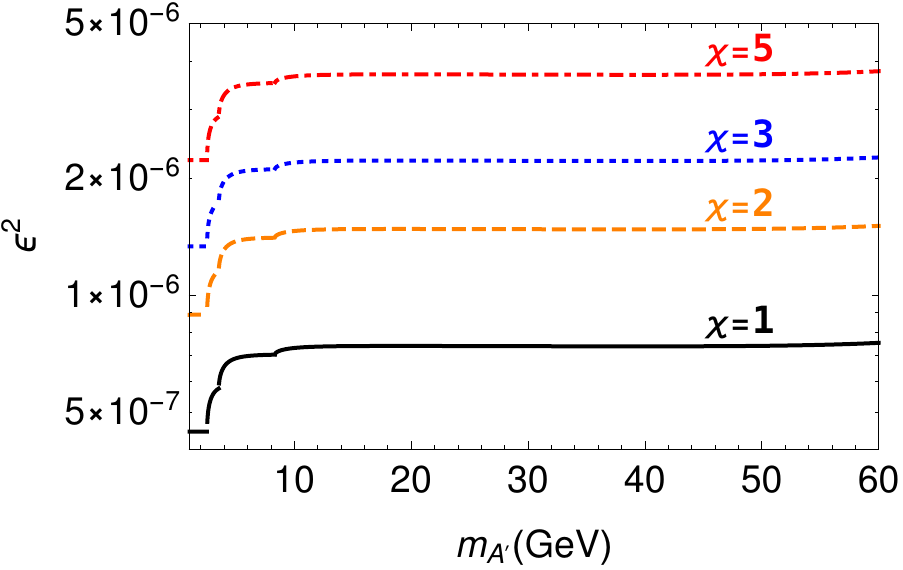}
\vspace*{1em}
\includegraphics[width=0.3\textwidth]{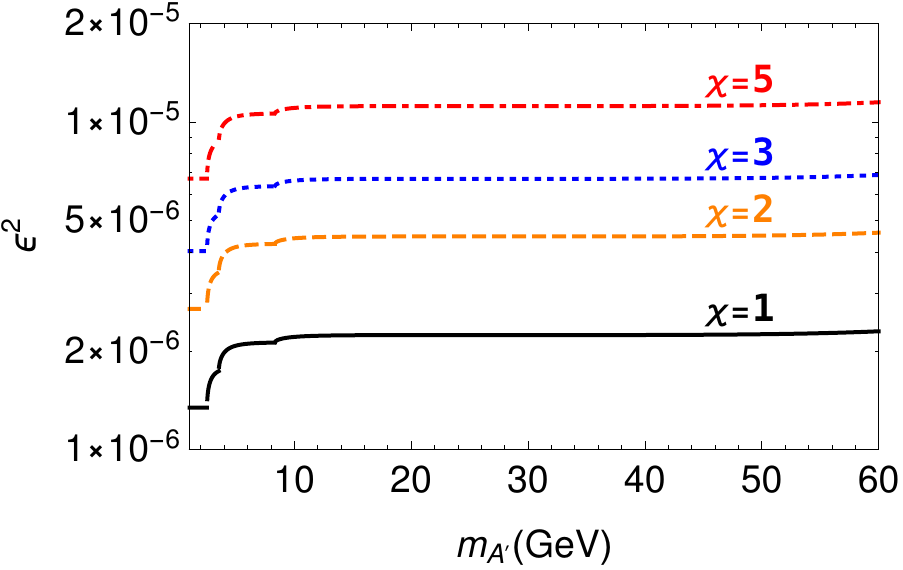}
\caption{Sensitivity on $\epsilon^2$ as functions of dark photon mass $m_{A^{\prime}}$ for a given $\chi$ for FCC-ee.
The figures from left to right are for $\sqrt{s} = 160$ GeV, 240 GeV and 350 GeV for one-year running, respectively.}
\label{fig5}
\end{figure}

\noindent
{\bf Conclusions}

In summary, we have studied the possibility of searching for dark photon interaction at a circular $e^+e^-$ collider through the process $e^+ e^-\to \gamma A^{\prime *} \to \gamma  \mu^+\mu^-$. There are two contributions in the SM, the intermediate $\gamma$ and intermediate $Z$. When the dark photon mass and also the center-of-mass frame energy are significantly away from the $Z$ boson mass, the dominate contribution to the SM background is from intermediate $\gamma$ interaction. The CEPC and FCC-ee $e^+e^-$ can provide sensitive constraints on the dark photon mixing parameter and dark photon mass. We find that the constraints on $\epsilon^2$ for dark photon mass in the few tens of GeV range, assuming that the $\mu^+\mu^-$ invariant mass can be measured to an accuracy of $0.5\%m_{A'}$,
can be better than $3\times 10^{-6}$ for the proposed CEPC with a ten-year running at 3$\sigma$ (statistic) level, and better than $2\times 10^{-6}$ for FCC-ee with even just one-year running at $\sqrt{s} = 240$ GeV, better than the LHC and other facilities can do in a similar dark photon mass. For FCC-ee, running at $\sqrt{s}=160$ GeV, the constraint can be even better. In the range of 20 GeV to 60 GeV for $m_{A'}$, the smallest $\sigma_{\mu\mu}$ is 100 MeV which is reachable at the CEPC and FCC-ee. With a smaller $\sigma_{\mu\mu}$, the sensitivity can be improved.

\begin{acknowledgments}

XGH was supported in part by MOE Academic Excellent Program (Grant No.~105R891505) and MOST of ROC (Grant No.~MOST104-2112-M-002-015-MY3), and in part by NSFC (Grant No. 11575111) of PRC. This work was also supported by Key Laboratory for Particle Physics, Astrophysics and Cosmology, Ministry of Education, and Shanghai Key Laboratory for Particle Physics and Cosmology (SKLPPC) (Grant No. 11DZ2260700).

\end{acknowledgments}

\end{document}